# Scaffolding Progress: How Structured Editors Shape Novice Errors When Transitioning from Blocks to Text

Majeed Kazemitabaar
Department of Computer Science
University of Toronto, Ontario, Canada
majeed@dgp.toronto.edu

Viktar Chyhir
Department of Computer Science
University of Toronto, Ontario, Canada
viktar@dgp.toronto.edu

David Weintrop
College of Education, College of Information Studies
University of Maryland, College Park, USA
weintrop@umd.edu

Tovi Grossman
Department of Computer Science
University of Toronto, Ontario, Canada
tovi@dgp.toronto.edu

## ABSTRACT

Transitioning from block-based programming to text-based programming environments can be challenging as it requires students to learn new programming language concepts. In this paper, we identify and classify the issues encountered when transitioning from block-based to text-based programming. In particular, we investigate differences that emerge in learners when using a structured editor compared to an unstructured editor. We followed 26 high school students (ages 12-16; *M*=14 years) as they transitioned from Scratch to Python in three phases: (i) learning Scratch, (ii) transitioning from Scratch to Python using either a structured or unstructured editor, and (iii) evaluating Python coding skills using an unstructured editor. We identify 27 distinct types of issues and show that learners who used a structured editor during the transition phase had 4.6x less syntax issues and 1.9x less data-type issues compared to those who did not. When these learners switched to an unstructured editor for evaluation, they kept a lower rate on data-type issues but faced 4x more syntax errors.

## 1 INTRODUCTION

Block-based programming environments (BBPEs), such as Scratch [22], are one of the prominent ways of introducing novices and young students to computer science and programming. These environments have been specifically designed to allow learners focus on "what can be programmed" rather than "how to program something", using two main approaches: (i) providing visual affordances and abstractions to editing code; and (ii) simplifying concepts through programming language design. Research has shown that BBPEs are effective at helping K-12 students develop computational thinking *concepts* such as sequence, loops, and conditionals, in addition to *practices* such as modularizing, debugging, and being iterative [12, 22, 27].

However, prior research has also identified several drawbacks to BBPEs such as failing to scale well with larger programs, being perceived as inauthentic and less powerful, and being more verbose and slower to author [28]. Additionally, in professional settings, BBPEs lack many of the benefits that come with text-based programming environments such as being easier to collaborate, supporting version control, and access to a vast ecosystem of libraries and tools designed to support programmers. As a consequence, novices may wish (or need) to transition from blocks to text as their education progresses.

Therefore, it is important to understand what challenges novices face when making this transition. To date, researchers have articulated a number of barriers, including difficulties associated with reading/parsing text-based code, memorizing syntax, typing in commands [15], and interpreting errors [5, 9]. Although these could potentially explain why this transition is difficult, the characteristics and frequency of the actual issues that novices face through this transition remains understudied.

Furthermore, in an attempt to ease this transition, hybrid environments like Frame-based editing [15] and CodeStruct [14] have incorporated structured editors that support learners by eliminating syntax errors and increasing readability through the display of visual cues. However, the effect of such structured editors on novice errors is not known. Particularly, its effect on errors while learners are making the transition or after the transition, when learners switch to unstructured editors. Therefore, in this paper, we seek to answer the following questions about issues that learners face in their *initial* text-based programming experience when transitioning from a BBPE:

- **RQ1:** what types of issues learners face? what are their characteristics, relative frequencies, and difficulty levels?
- **RQ2:** how do the frequencies of issues differ between a structured and unstructured editor?
- **RQ3:** how do the frequencies of issues change when transitioning from a structured to an unstructured editor?

To answer these research questions, we examine data from a previous study [14] conducted with 26 novices in grades 7-11 (ages 12-16; *M*=14 years) with no prior text-based programming experience that transitioned from Scratch to Python in 11 90-minute sessions. The sessions were video recorded and thematically analyzed. The result of this analysis deepens our understanding of the issues novices face in transitioning from block-based to text-based programming. Further, this analysis reveals ways in which, with the right supports, novices can resolve their issues. Our findings identify 27 types of issues in three major categories: *syntax* (48%), *data-types* (32%), and *semantics* (20%). Learners who used a structured editor during the transition phase had 4.6x less syntax errors and 1.9x less data-type issues compared to those that didn't. In addition, the structured editor group had 4x more syntax errors but similar data-type issues when transitioned into using an unstructured editor during the post-transition phase. Collectively, this work contributes to the growing literature on ways to support novices in transitioning from block-based to text-based programming, specifically as it relates to the ways that structured editors can support this transition.

## 2  RELATED WORK

### 2.1 Transitioning from Blocks to Text

A question of increasing consequence about BBPEs is whether the approach effectively prepares learners for text-based programming environments. Early work on graphical (but not block-based) programming [24] found little evidence of successful transfer of basic programming tasks in novices when moving to text-based programming languages despite prior programming success with graphical tools. Similar outcomes have been documented using contemporary block-based environments, with case studies documenting challenges associated with the transition [9, 13, 21]. On the contrary, there are studies showing that programming skill and knowledge gains made in block-based tools *do* prepare learners for later text-based programming. Armoni et al. [2] conducted a longitudinal study to investigate this topic. The resulting analysis reported little quantitative difference in their performances on assessments but did identify some specific content areas where the students with earlier Scratch experience outperformed their peers (e.g., iterating programming constructs). Additionally, the authors found that students with prior Scratch experience reported higher levels of self-efficacy and motivation to learn to program. Grover et al. [12] showed that learners performed well on conceptual multiple-choice questions posed in text-based languages after working through a Scratch-based curriculum. These studies suggest that block-based tools can help learners in future text-based contexts, especially on conceptual knowledge. However, little is known about how they prepare code composition skills as learners initially transition to text-based tools.

### 2.2 Introductory Programming Challenges

Prior work in exploring common misconceptions and challenges in introductory programming such as the work done by Hristova et al. [13] has identified 20 common programming errors in Java after surveying college professors, teaching assistants, and students in Computer Science. They combined the data into three major categories: *syntax*, *semantics*, and *logic*. Afterwards, Altadmri and Brown [1] used these 20 programming errors to automatically analyze a large data set of user-compiled Java code from Blackbox [6]. They prioritized errors by frequency and explored how the trends changed over time for each user. Furthermore, although the issues that are specific to transitioning from blocks to text have not been formally investigated in the literature, Kölling et al. [15] mentions 13 fundamental problems related to the transition and argue that many of the problems amplify each other when encountered in conjunction. Yet, the exact characteristics of the encountered issues, their frequency, and difficulty are still understudied.

### 2.3 Easing the Blocks-to-Text Transition

To overcome the challenges associated with transitioning from block-based to text-based programming, three distinct approaches have been identified by Lin & Weintrop [17]: one-way transition, dual-modality, and hybrid. *One-way transition environments* allow novices to view or export the text-based version of a block-based code that they authored. *Dual-modality environments* such as MakeCode [3] and Pencil Code [4] allow learners to author their programs in either block-based or text-based forms and support them in moving back-and-forth between the two. *Hybrid environments* like Frame-based editing [15], TouchDevelop [26], and CodeStruct [14] combine features of block-based and text-based programming into a single interface. These environments are usually structured editors and edits are performed directly to the Abstract Syntax Tree (AST) to always maintain a valid structure. Prior work has shown that transitioning to these environments can reduce task completion time and required instructor assistance when compared to transitioning directly to text [17, 22]. However, the effect of structured editors on the types of issues, their characteristics, and frequency are not well investigated.

Therefore, this work is differentiated from prior work by reporting the results of a unique study in which, (i) participants experience text-based programming for the first time, (ii) learners carry knowledge and skill from a BBPE, (iii) code composition skills are evaluated in addition to conceptual understanding, and (iv) learners go through the transition independently using their programming tools and documentations instead of relying on formal lectures.

## 3 METHODS & PARTICIPANTS

### 3.1. Data Set

Our data set is based on a prior two-week long study, in which students transitioned from Scratch to Python [14]. In this paper, we reuse the same data set and complement their prior analysis of task performance measures (correctness score, completion time, and number of help requests), with a more thorough and detailed analysis of the types of transition issues that occurred, their characteristics, and how they were resolved.

### 3.2. Study Design

The study was comprised of eleven 90-minute sessions, conducted over the Google Meet platform, and broken down into three phases. The first phase had participants use Scratch and included a mixture of lectures and programming tasks. The second phase had participants transition independently from block-based to text-based programming in four sessions. Half of the learners made this transition using a structured editor called CodeStruct [14]. The other half of the learners transitioned directly from Scratch to Python using Replit which is unstructured. Finally, code composition skills and conceptual understanding were evaluated in the last two sessions.

### 3.3 Participants

Participants were recruited through a local school and an after-school program in a North American city. The study consisted of 26 students (14 female/12 male) ages 12-16 (*M*=14.1; *SD*=1.2) and none had prior text-based programming experience. Parent/guardian consent was obtained before the first session and each participant received a $50 gift card at the end.

### 3.4 Instructional Sequence

To begin the study, all youth in the study went through a 4-session introductory sequence with Scratch. Each session included one or two lectures on the basic programming concepts followed by 3-6 programming activities. These sessions covered the basics of Scratch programming (e.g., sequence), variables, conditionals, iterative logic, and arrays.

In the second phase of the study, participants were asked to complete 18 programming tasks in which they had to convert a given screenshot of a Scratch code to Python and a further 15 tasks in which they had to write a Python program based on the given description and sample outputs. The programming tasks were inspired by prior studies [29] and evaluated common programming misconceptions [25]. To provide additional assistance, all learners were allowed to ask questions when they were stuck. Additionally, they were given a documentation that included all the Python sections of a novice-friendly tutorial from w3schools.com. Students completed these programming tasks using either CodeStruct or Replit. Replit includes many functionalities found in professional code editors such as code completion, syntax highlighting, and static code analysis. Additionally, CodeStruct includes (i) a structured code editor that eliminates most syntax errors, provides code completion, active data-type checking, and immediate fix suggestions, (ii) a context-aware toolbox with basic code snippets and on-hover learning material that supports point-and-click code insertion at the caret, and (iii) various visual cues drawn from BBPEs to help novices know how and where commands can be used, such as holes for empty arguments and highlighted code blocks. A detailed specification of all its features can be found in the Design and Evaluation of CodeStruct [14].

The final phase of the study asked students to author additional Python programs designed to evaluate (i) Python code composition skills, (ii) python-specific concepts (e.g., for loops), and (iii) Scratch to Python conceptual differences (e.g., while loop conditions and list indexing). Both conditions in this phase used Replit. Content assessments and attitudinal surveys were administered but not analyzed for the present work.

### 3.5 Data Collection & Analysis

A variety of measures and observations were collected throughout the study, including video recordings, online form data, content assessments, and log data on student interactions with the programming tools and documentation. For video recordings, all participants were asked to share their entire screen which was recorded. Instead of analyzing learner's task performance measures such as correctness score, completion time, and help requests that are reported in the original CodeStruct paper [14], this work extracts the characteristics of the encountered issues and tracks how they were fixed (as a measure of difficulty). For that, we used a thematic qualitative approach to analyze 230 hours of screen recordings from six sessions of the 26 students [23] in which they either used CodeStruct or Replit. Videos were initially broken down into multiple labeled segments corresponding each of the programming tasks using the ATLAS.ti Qualitative Data Analysis Software. To identify the characteristics of the issues and ways they were fixed, we pursued an iterative coding scheme. First, two researchers independently extracted common issues and how they were fixed in 10% of the videos (all videos from four randomly selected participants) to create an initial codebook. Afterwards, the two researchers met to discuss and co-interpret common themes and merge the codebook. Furthermore, to determine how issues were fixed and measure their difficulty, five tags were added to the codebook for each identified issue: (i) fixed after receiving help, (ii) fixed after referring to the documentation, (iii) fixed using the tool, (iv) fixed by the learner themselves, and (v) remained unfixed. The researchers then independently tested the codebook on 13 questions from two participants and reached an 84% agreement on the coded issues (both in terms of identifying the same issues and using similar tags). After merging conflicts and finalizing the codebook, one researcher analyzed videos from all participants.

## 4 RESULTS

### 4.1 Transition Issues by Editor Type

In this section we answer our first and second research questions that ask about the types of issues learners face in their initial transition from Scratch to Python along with their relative frequencies, difficulty levels, and how they differ between a structured and unstructured editor. For that we report aggregated results over the four sessions in the transition phase, broken down by group (abbreviated into CS for CodeStruct, and R for Replit). Additionally, for issues that caused an error, we report its difficulty by calculating how often it was left unfixed or required help to be fixed, divided by how often it occurred.

Overall, our analysis identifies 27 types of recurring issues (22 errors, and five misconceptions) that learners faced in their initial transition from Scratch to Python (Figure 1). Errors are distinguished with an (E) and caused an error message to be displayed, while misconceptions (M) did not. These 27 issues can be broken down into three categories: *data-types*, *syntax*, and *semantics* that are discussed in the following three subsections.

*4.1.1 Syntax Issues by Editor Type*

The most frequent category of issues was syntax issues, including incorrect usage of *punctuation*, *operators*, *string quotes*, *function calls*, *variables*, *lists*, *importing modules*, and *conditional statements*. These issues mostly stem from the syntax rules used in Python. See Figure 1 (b-i) for more information about each of the issues, their relative frequencies, and how they were fixed.

Overall, we documented 348 (48%) syntax issues (CS: 62, R: 286), which occurred about five times less in the CodeStruct group. This is mainly due to the structured nature of CodeStruct that almost eliminates syntax issues on *missing or misplaced colons*, *mismatched parenthesis*, *incorrect operator syntax*, *incorrect function calls*, and *using elif without a condition*. The most frequent types of syntax issues were: (i) *missing or incorrect usage of string quotes* (CS: 10, R: 42, difficulty: 0.27), (ii) *incorrect syntax of comparator operators* such as using = instead of == to check for equality (CS: 0, R: 34, difficulty: 0.35), (iii) *incorrect string concatenation* in Python such as typing join similar to Scratch instead of using the plus operator (CS: 11, R: 18, difficulty: 0.34), and (iv) *incorrect variable assignment* (CS: 2, R: 15, difficulty: 0.35). We also identified a few syntax misconceptions that showed a misunderstanding or a lack of knowledge about Python's syntax rules: (i) not knowing how to initialize an empty list using `items = []` and instead, using `items = [0]` and clearing it with `items.pop()`, or (ii) not knowing how to use commas to initialize a list with multiple values (e.g. `items = [0, 1, 2, 3]`), and (iii) importing the random module for each usage of the `randint()` function (Figure 2b).

*4.1.2 Data-Type Issues by Editor Type*

The second major category of issues was data-types with 229 (32%) documented occurrences (CS: 79, R: 150), which occurred about two times less in the CodeStruct group. CodeStruct's structured editor updates the data type of all variables and literals in the AST as code is being edited. Therefore, it can highlight data-type mismatches with fix suggestions inside the editor even before code execution.

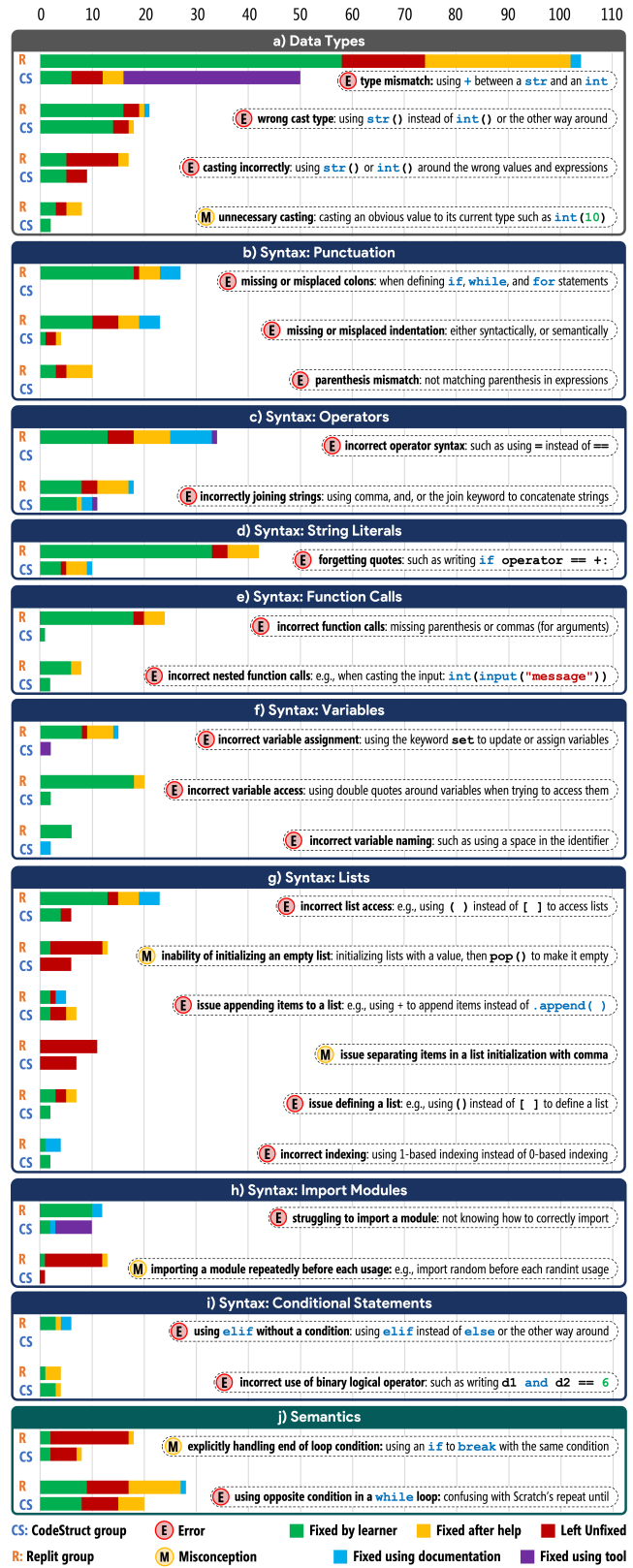

Figure 1: Scratch to Python transition issues

```
while not (ask == "stop"):
    # do something with ask

    if ask == "stop":
        break
```
**a.** Explicitly breaking out of a while loop using a break statement with the same condition

```
import random
first = random.randint(1, 10)
import random
second = random.randint(1, 10)

# do something with first and second
```
**b.** Importing a module that is already imported before using it again.

**Figure 2: Common transition misconceptions**

The most common type of issue in this category was *type mismatch errors* which happened when unmatching types were used with an operator (either arithmetic or comparator). Forgetting to cast types resulted into a type mismatch error after code execution. This issue had 154 occurrences (CS: 50, R: 104, difficulty: 0.35). There were two other major patterns related to data-types: (i) *casting with an incorrect type* (CS: 18, R: 21, difficulty: 0.2) which shows that learners knew that they had to use some type of casting to fix their type mismatch error but did not know the correct casting type, and (ii) *unnecessary casting* (CS: 9, R: 17) which was a misconception of not knowing that casting a value to its current type did not have any effect.

*4.1.3 Semantic Issues by Editor Type*

The final major type of issue was *semantics* with 74 occurrences (CS: 28, R: 46), mostly occurring in the context of conditional loops. These issues attributed to 17 help requests and were one of the most difficult issues to be fixed by the learners themselves (only 28% from both groups). We documented two types of recurring patterns: (i) *using the opposite condition in a while loop* (CS: 20, R: 26, difficulty: 0.62), and (ii) *explicitly handling the end of a conditional loop using an if statement* to break from a loop (CS: 8, R: 18) which is portrayed in Figure 2a. The first issue is probably due to the transition from Scratch to Python as the **while** loop in Python repeats when the condition is true, while the **repeat until** block in Scratch repeats when the condition is false. The second issue that does not cause errors in the program, captures a misconception that is probably also stemmed from the same change in paradigm.

## 4.2 Post-Transition Issues by Editor Type

Our third research question asks about how the frequencies of issues change when transitioning from a structured to an unstructured editor. To answer this question, we focus on comparing the results in the post-transition phase in which both groups used Replit, an unstructured editor and were evaluated for their code composition skills. This was a second transition for learners in the CodeStruct group that used a structured editor during the transition phase, and overall, they encountered more issues in the post-transition phase (CS: 184, R: 134). In the last phase, most of the differences between the two conditions came from syntax issues, specifically, *punctuation* issues (such as missing or misplaced colons and indentations) that occurred nearly four times more in the CodeStruct group (CS: 80, R: 21).

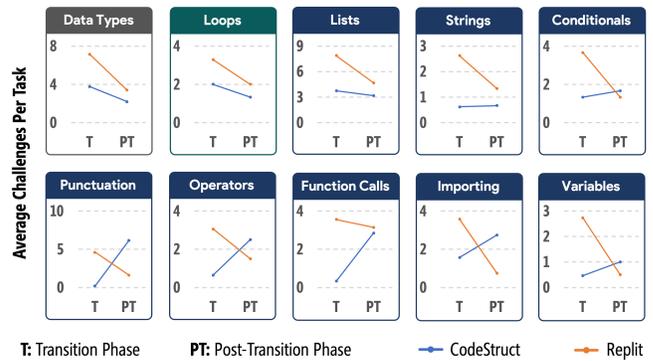

**Figure 3: Comparing average issues per task in the transition phase with the post-transition phase**

To report how the types of issues and their frequencies changed when learners transitioned from a structured editor to an unstructured editor, we compare the normalized count of issues in each phase based on the number of programming tasks in which they occurred in. Figure 3 illustrates a summary of this metric for each category of issues. For several syntax issues, the normalized *issues per task* increased considerably in the CodeStruct group (Figure 3 Bottom), while decreased in the Replit group: punctuation (CS: 0.22 → 6.15, R: 4.61 → 1.61), operators (CS: 0.65 → 2.5, R: 3.05 → 1.5) and importing modules (CS: 1.57 → 2.75, R: 3.57 → 0.75). However, we did not observe a similar pattern with variables, strings, lists, and conditionals. This is probably because CodeStruct automatically handled punctuation while requiring users to explicitly handle most of the syntax for variables, strings, lists, and conditionals.

Moreover, there was a slight drop in both conditions for semantic issues (CS: 2 → 1.33, R: 3.28 → 2) and data types (CS: 3.76 → 2.2, R: 7.14 → 3.4). This is a promising result: showing that learners from the CodeStruct group who received immediate feedback on type mismatch issues were able to maintain a lower rate on data-type issues in the post-transition phase, while no longer having access to CodeStruct.

## 5 DISCUSSION

Our results complement and extend prior analysis of CodeStruct's features and task performance metrics, including correctness scores and completion time [14]. In this paper, we identified 27 types of issues that learners face when transitioning from Scratch to Python, including five misconceptions and 22 errors. Using a structured editor like CodeStruct during the transition can almost completely prevent six syntax-related issues and substantially reduce the occurrence of 12 other issues. However, this study found that learners who transitioned from a structured editor to an unstructured editor tended to regress on the issues that were prevented in the structured editor, while maintaining a low error rate on other issues such as data types that were simply identified by the structured editor.

In the sections below, we discuss the impact of our newly identified results on tool design and curriculum development.

## 5.1 Skill Development with Structured Editors

Our findings from RQ2 and RQ3 reveal that structured editors like CodeStruct are more effective when learners are actively involved in writing code. For example, CodeStruct automatically handled most syntax-related issues, such as *punctuation*, *operators*, and *function calls*. However, for *type mismatch errors* and *missing double quotes* for strings, CodeStruct only highlighted the incorrect code. The user could then hover over it, read a description of the issue, and use one of the suggested options to fix it. This active approach provides learners the opportunity to learn from the issue identified by the editor while still maintaining a low error rate in the post-transition phase. In contrast, the passive approach caused learners to regress. Future work on structured editors and programming environments for novices could incorporate more active approaches for handling *colons*, *indentations*, *operators*, *parentheses*, *function calls*, *imports*, and *variables* to support proper skill development.

## 5.2 Gradual Transition from Blocks to Text

Another approach to ease the transition from blocks to text, is to develop programming environments that gradually introduce new concepts (similar to Blocks-to-CAD [16]). For example, a Scratch-to-Python transition environment could start in blocks and gradually introduce new concepts related to Python like *data-types*, *imports*, and *conditional loops*. This incremental approach allows learners to practice and learn about the semantic principles of Python in a block-based environment and become familiar with the new syntax. When the learner is ready, the tool could switch to a text-based editor for practicing syntax. Our list of identified issues, their difficulty levels, and frequency can help inform how best to incrementally introduce new concepts and break down the transition steps.

## 5.3 Leveraging Prior Scratch Knowledge

Given the ubiquity of introductory environments like Scratch, effectively leveraging on learners' prior programming knowledge and experiences is important. Structured editors seeking to aid the transition beyond block-based programming, should draw on these prior experiences. For example, new topics such as *data-types*, *conditional loops*, *operators*, *string literals*, *joining strings*, *function calls*, and *nested function calls* can be linked to similar concepts in Scratch to make them less overwhelming. This could be done in static, or interactive ways. For instance, subgoal labeled, worked examples [19] that display the Python and Scratch modality of the same code side-by-side, can provide a one-to-one mapping between the subgoals of each modality. Additionally, interactive versions of these worked examples could be developed that simultaneously trace through both modalities and display the internal state of the program, in addition to its output. These interactive worked examples can better demonstrate the differences between the two modalities, particularly on topics related to *data-types*, *conditionals*, and *conditional loops*.

## 5.4 Detect Issues with Large Language Models

To ensure that students have the opportunity to actively learn and receive valuable feedback, it is important to allow them to make mistakes, and then be able to identify and locate the errors that they make. Structured editors are able to detect such fine-grained issues like *missing double quotes*, *misplaced parenthesis*, and *colons* much easily. However, for many other types of issues presented in this paper, an active learning approach requires using unstructured editors to allow users to make mistakes. These unstructured editors use code analyzers to identify errors in code, but they are limited in their ability to detect fine-grained errors and have low accuracy in locating them. Additionally, they do not work well when multiple errors are present and cannot detect data-type issues in dynamically typed languages. Recent advancements in Deep Learning have led to the development of models like OpenAI Codex [7], which can generate code from descriptions, explain code, and fix errors [18]. These models can be fine-tuned on samples of errors and misconceptions identified in this paper. The fine-tuned models can accurately annotate problematic pieces of code by localizing their issues and including suggestions on how to fix them. Future work may involve in using AI models like these to assist novice programmers.

## 5.5 Limitations and Future Work

In the context of transitioning from Scratch to Python, our identified issues cover most topics that are common between Scratch and Python, with the except function definitions. Future work should explore function definitions and working with arguments in more detail. Our results are limited by the number of participants in the data set, the overall complexity of tasks, and number of tasks used for each topic. To improve the accuracy of our results, future work could collect keystroke data from learners and use this data to measure the frequency of errors and misconceptions programmatically. This would be an improvement over our current method of conducting a thematic video analysis. Additionally, although transitioning from Scratch to Python is one of the most popular paths from blocks to text, it does not account for changes in programming language. Therefore, future work could study how students transition from a block-based Python like EduBlocks [10] to text-based Python.

## 6 CONCLUSION

In this paper, we present the top recurring errors and misconceptions that novice programmers face when transitioning from a block-based programming environment like Scratch to a text-based programming environment like Python. We identified 27 types of issues and their frequencies, measured their difficulty, and compared how they differed when learners transitioned using either a structured or unstructured editor. We also report how switching from a structured editor to an unstructured editor changes the frequency of issues. Our findings provide insight into the challenges that students face and can inform the design of future assistance mechanisms, intermediary tools, and curriculum development.